\documentclass[11pt]{article}

\usepackage[final]{acl}
\usepackage{amsmath}
\usepackage{amssymb}
\usepackage{times}
\usepackage{latexsym}
\usepackage[T1]{fontenc}
\usepackage[utf8]{inputenc}
\usepackage{microtype}
\usepackage{inconsolata}
\usepackage{graphicx}
\usepackage{booktabs}
\usepackage{amsmath}
\usepackage{tikz}
\usepackage{xcolor}
\usetikzlibrary{arrows.meta,positioning}

\title{UZH-CL at ArA-DF 2026: Prompt-Tuned Foundation Models and\\ Track-Adaptive Score Fusion for Arabic Speech Deepfake Detection}

\author{
  \textbf{Aref Farhadipour}$^{1}$,
  \textbf{Teodora Vukovic}$^{1}$,
  \textbf{Petr Motlicek}$^{2,3}$ \\
  $^{1}$Department of Computational Linguistics, University of Zurich, Switzerland \\
  $^{2}$Idiap Research Institute, Switzerland \qquad
  $^{3}$Brno University of Technology, Czech Republic \\
  \texttt{aref.farhadipour@uzh.ch}
}

\begin{document}
\maketitle

\begin{abstract}
Detecting synthetic and voice-converted speech remains difficult for low-resource languages with dialectal diversity, where systems must generalize across regional dialects and unseen acoustic channels. We present the UZH-CL submission to the ArA-DF 2026 Shared Task on Arabic speech deepfake detection, covering Track~1 (dialect generalization) and Track~2 (acoustic robustness). We freeze a W2V-BERT-2.0 backbone and adapt it with \emph{Wavelet Prompt Tuning}, updating under 1\% of parameters, and aggregate multi-layer representations with cross-layer attention and a general attentive-statistics pooling head rather than a specialized graph backend. Complementary detectors are obtained by varying adaptation strategy, training data, augmentation, and encoder family. We find that the two shift types require different fusion regimes: a broad multi-window ensemble for dialect generalization, and a compact, channel-matched, center-crop ensemble for acoustic robustness. Official evaluation yields 1.96\% EER on Track~1 (6th place) and 1.04\% EER on Track~2 (3rd place), corresponding to 87\% and 96\% relative reductions over the XLS-R+AASIST baselines.
\end{abstract}

\section{Introduction}

Neural text-to-speech (TTS) and voice conversion (VC) can generate synthetic Arabic speech that can be used for media forgery, impersonation, and fraud. Detecting such speech is more difficult than in English-centric settings, because Arabic includes Modern Standard Arabic (MSA) and regional dialects that differ in phonology, morphology, and prosody, and because recordings may include codec compression, background noise, room reverberation, and transmission filtering. The ArA-DF 2026 Shared Task evaluates these conditions in two tracks: Track~1 (\emph{dialect generalization}) and Track~2 (\emph{acoustic robustness}).

We address both tracks with parameter-efficient adaptation of a speech foundation model and with score fusion that depends on the type of shift. Fine-tuning a large speech model on a small in-domain set can overfit and overwrite pre-trained features. We freeze W2V-BERT-2.0~\cite{barrault2023seamless} and train \emph{Wavelet Prompt Tuning} (WPT)~\cite{farhadipour2026wpt}. WPT prepends standard prompts and Haar-wavelet prompts to the input of each transformer layer, so less than 1\% of the parameters are updated.

Many anti-spoofing systems, including the official baseline, take features from a single SSL layer and classify them with AASIST~\cite{jung2022aasist}. We instead combine all transformer layers with cross-layer attention and a multi-head attentive-statistics pooling head. Lower layers tend to carry spectral artifacts; higher layers carry more linguistic content. Using all layers lets us keep a general pooling head rather than a graph backend such as AASIST.

We also train detectors that differ in adaptation, training data, and augmentation, and that use other encoders besides W2V-BERT: XLS-R~300M, Dasheng~1.2B, and the Qwen3-Omni audio encoder. Score fusion is set separately for each track. Track~1 benefits from overlapping windows and a larger set of less correlated detectors; Track~2 benefits from a short center crop and a smaller set of channel-augmented models.


\section{Background and Task Setup}
\label{sec:task}

\paragraph{Related work.}
Speech anti-spoofing systems typically combine a self-supervised learning (SSL) frontend (such as wav2vec\,2.0~\cite{baevski2020wav2vec2} or XLS-R~\cite{babu2021xlsr}) with graph neural network backends such as AASIST~\cite{jung2022aasist,tak2022ssleer}. However, full fine-tuning of large SSL backbones is computationally expensive and prone to overfitting on small target datasets. Parameter-efficient fine-tuning (PEFT) methods, including prefix tuning~\cite{li2021prefix} and visual prompt tuning~\cite{jia2022vpt}, address this by inserting small trainable prompt vectors into transformer blocks while freezing backbone weights. WPT~\cite{farhadipour2026wpt} extends prompt tuning to speech signals by incorporating multi-resolution frequency subbands via discrete wavelet transforms. 

\paragraph{Task and corpus setup.}
The ArA-DF 2026 Shared Task evaluates binary classification of 16\,kHz single-channel audio recordings into \emph{bonafide} (authentic) or \emph{spoofed} (synthetic or voice-converted) speech. The dataset provides 22.5k labeled training utterances and 21k labeled development utterances covering MSA and regional Arabic dialects. Evaluation is divided into two tracks, each featuring an unlabeled development-test set (16.0k files for Track~1; 14.2k for Track~2) used during challenge development, and an unlabeled final-test set (144.2k files for Track~1; 127.7k for Track~2) for official benchmark ranking. The evaluation metric is Equal Error Rate (EER). The official baseline system (XLS-R~300M + AASIST) yields final-test EERs of 14.54\% on Track~1 and 27.11\% on Track~2. Ground-truth test labels were released post-evaluation, enabling the per-detector diagnostics in Section~\ref{sec:permodel}.

\section{System Architecture}
\label{sec:system}

\begin{figure}[t]
\centering
\includegraphics[width=1.0\columnwidth]{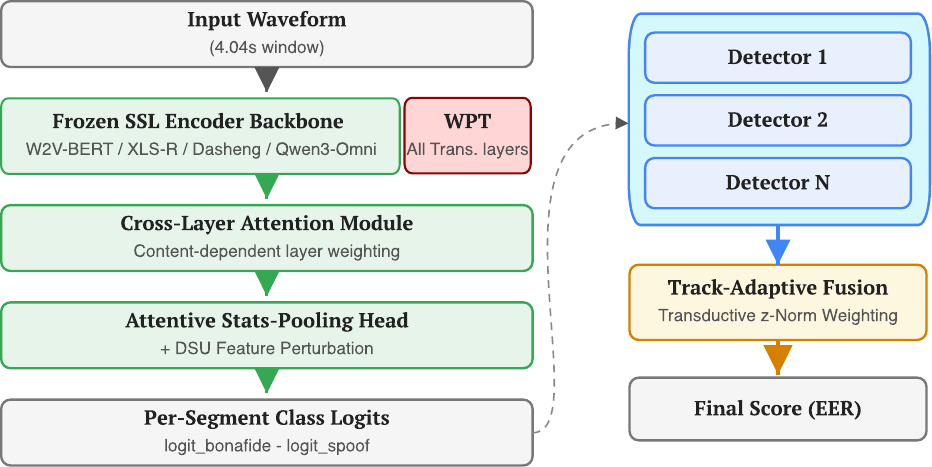}
\caption{System architecture: per-detector scoring followed by track-adaptive late fusion.}
\label{fig:system}
\end{figure}

Figure~\ref{fig:system} shows the two-stage pipeline. Each detector adapts a frozen SSL backbone (WPT or FT$N$), aggregates layers with cross-layer attention and attentive-statistics pooling, and outputs a scalar score. Late fusion then $z$-normalizes and weights those scores: multi-window averaging on Track~1 and a center crop on Track~2.

\begin{table*}[t]
\centering
\small
\resizebox{0.93\textwidth}{!}{%
\setlength{\tabcolsep}{4pt}
\begin{tabular}{l l l l l c c c c c}
\toprule
 & & & & & \textbf{dev} & \multicolumn{2}{c}{\textbf{Track 1 EER (\%)}} & \multicolumn{2}{c}{\textbf{Track 2 EER (\%)}} \\
\cmidrule(lr){6-6}\cmidrule(lr){7-8}\cmidrule(lr){9-10}
\textbf{ID} & \textbf{Encoder} & \textbf{Adapt.} & \textbf{Training Data} & \textbf{Augmentation} & in-dist & dev-test & final & dev-test & final \\
\midrule
T1   & W2V-BERT-2.0 & WPT   & ArA-DF (22.5k)       & DSU            & 0.04 & 9.06 & 9.34 & 16.81 & 17.32 \\
T2   & W2V-BERT-2.0 & WPT   & ArA-DF (22.5k)       & DSU + RB + MR  & 0.22 & 2.86 & 3.01 & 1.99  & 1.83  \\
S123 & W2V-BERT-2.0 & WPT   & ArA-DF (22.5k)       & DSU + RB       & 0.23 & 2.76 & 3.15 & 1.80  & 1.76  \\
S456 & W2V-BERT-2.0 & WPT   & ArA-DF (22.5k)       & DSU + RB       & 0.19 & 2.98 & 3.06 & 2.08  & 1.98  \\
C123 & W2V-BERT-2.0 & WPT   & Arabic ($\approx$100k) & DSU + RB + MR  & 0.58 & 3.10 & 3.31 & 2.73  & 2.68  \\
C456 & W2V-BERT-2.0 & WPT   & Arabic ($\approx$100k) & DSU + RB + MR  & 0.22 & 3.32 & 3.53 & 1.63  & 1.58  \\
FTBB & W2V-BERT-2.0 & FT16  & Arabic ($\approx$100k) & DSU + RB + MR  & 0.28 & 3.34 & 3.37 & 1.57  & 1.46  \\
FTX  & W2V-BERT-2.0 & FT16  & Multi ($\approx$264k)  & DSU + RB + MR  & 0.43 & 3.80 & 3.66 & 1.76  & 1.59  \\
FTXA & W2V-BERT-2.0 & FT12  & Multi ($\approx$264k)  & DSU + RB + MR  & 0.22 & 3.23 & 3.29 & 1.45  & 1.39  \\
MEGA & W2V-BERT-2.0 & WPT   & Multi ($\approx$264k)  & DSU + RB + MR  & 0.45 & 3.61 & 3.56 & 2.14  & 2.08  \\
DIA  & W2V-BERT-2.0 & FT16  & Arabic ($\approx$100k) & Speed + DSU    & 0.27 & 6.73 & 6.53 & 11.68 & 11.63 \\
CDEV & W2V-BERT-2.0 & WPT   & ArA-DF + dev         & DSU + RB + MR  & 0.62 & 4.54 & 4.66 & 2.11  & 2.13  \\
XLSR & XLS-R 300M   & WPT   & Arabic ($\approx$100k) & DSU + RB + MR  & 1.15 & 5.64 & 5.84 & 5.50  & 5.49  \\
DSH  & Dasheng 1.2B & WPT  & Arabic ($\approx$100k) & DSU + RB + MR  & 1.37 & 11.35 & 12.10 & 3.42 & 3.60  \\
QWN  & Qwen3-Omni   & WPT   & Arabic ($\approx$100k) & DSU + RB + MR  & 1.07 & 9.15 & 8.88 & 3.03  & 3.03  \\
\midrule
\multicolumn{5}{l}{\textit{Best single model}}                           & --   & 2.76 & 3.01 & 1.45  & 1.39  \\
\multicolumn{5}{l}{\textbf{Track-adaptive fusion (ours)}}                & --   & \textbf{1.83} & \textbf{1.96} & \textbf{1.17} & \textbf{1.04} \\
\bottomrule
\end{tabular}}
\caption{Constituent base detector specifications and per-split EER (\%). Adaptation: WPT (frozen backbone) or FT$N$ (top-$N$ layers fine-tuned). Training data: ArA-DF = official training set (22.5k clips); Arabic = expanded Arabic pool ($\approx$100k clips); Multi = multilingual pool ($\approx$264k clips). Augmentation: RB = RawBoost, MR = MUSAN/RIR noise/reverberation. Column ``dev'' reflects in-distribution validation EER (single 4\,s crop). Track 1 and Track 2 test columns report scores under track-specific segment aggregation (multi-window vs.\ center-crop).}
\label{tab:detectors}
\end{table*}

\subsection{WPT Core Detector}
\label{sec:core}
The primary core detector uses a frozen W2V-BERT-2.0 backbone~\cite{barrault2023seamless} consisting of $L=24$ Conformer layers with hidden dimension $d=1024$. At layer $\ell$, given input frame representations $H_{\ell-1}\in\mathbb{R}^{T\times d}$, WPT prepends $n_p=6$ standard prompt vectors $P_\ell\in\mathbb{R}^{n_p\times d}$ and $n_w=4$ wavelet prompt vectors derived from trainable parameters $Q_\ell\in\mathbb{R}^{n_w\times d}$ via a discrete Haar wavelet transform block $\Psi$:
\begin{align}
(A_\ell, D_\ell) &= \mathrm{DWT}_{\text{haar}}(Q_\ell), \nonumber\\
\Psi(Q_\ell) &= \mathrm{Proj}\big([A_\ell \,;\, D_\ell]\big),\\
H'_{\ell-1} &= [\,\Psi(Q_\ell)\,;\, P_\ell\,;\, H_{\ell-1}\,],
\end{align}
where $A_\ell, D_\ell \in \mathbb{R}^{n_w\times(d/2)}$ denote low-frequency approximation and high-frequency detail subbands, and $\mathrm{Proj}$ projects the concatenated subbands back to dimension $d$. The modified sequence $H'_{\ell-1}$ passes through transformer block $f_\ell$, after which prompt representations are removed prior to layer $\ell+1$: $H_\ell = f_\ell(H'_{\ell-1})_{[\,n_w+n_p:\,]}$. During training, backbone weights $f_\ell$ remain frozen; only prompt parameters $\{P_\ell, Q_\ell, \Psi\}$, cross-layer modules, and downstream pooling heads are optimized ($<1\%$ of parameters).

To aggregate representations across depth, a 4-head \emph{cross-layer attention} module computes content-dependent weights over layer outputs $\{H_\ell\}_{\ell=1}^{24}$, reflecting that spectral artifacts and high-level linguistic content concentrate at different encoder layers. The aggregated representation is processed by a multi-head attentive-statistics pooling head~\cite{okabe2018attentive,desplanques2020ecapa} (4 heads, 128-d bottleneck, 256-d embedding) to produce binary class logits. Feature perturbation via Domain-Shift-with-Uncertainty (DSU)~\cite{li2022dsu} is applied inside the pooling head during training alongside a class-weighted cross-entropy loss (weight 3.0 on bonafide). The scalar detection score for an input $x$ is defined as $g(x) = \mathrm{logit}_{\text{bonafide}}(x) - \mathrm{logit}_{\text{spoof}}(x)$.

\subsection{Detector Ensemble Diversification}
\label{sec:diversity}
To obtain complementary error patterns across constituent models, we diversify detectors along four operational axes (Table~\ref{tab:detectors}):
\begin{enumerate}
    \setlength{\itemsep}{1pt}\setlength{\parsep}{0pt}\setlength{\parskip}{1pt}
    \item \textbf{Adaptation strategy:} We contrast frozen WPT against \emph{partial backbone fine-tuning} (FT$N$), unfreezing the top $N\in\{12,16\}$ layers at $10^{-5}$ from a pre-trained WPT model. FTBB, FTX, and FTXA in Table~\ref{tab:detectors} are detector IDs using FT16 or FT12.
    \item \textbf{Encoder family:} Beyond W2V-BERT, we evaluate XLS-R~300M~\cite{babu2021xlsr} (waveform frontend), Dasheng-1.2B~\cite{dinkel2024dasheng} (spectrogram ViT), and the audio encoder extracted from Qwen3-Omni-30B-A3B~\cite{qwen3omni2025} (32 layers, mel-spectrogram frontend).
    \item \textbf{Training data composition:} We evaluate three training pools: official ArA-DF; an expanded \emph{Arabic pool} ($\approx$100k clips) incorporating ArFake~\cite{elsetohy2025arfake}, AFAD-MSA~\cite{issa2026afad}, an Arabic deepfake corpus~\cite{arabic_audio_deepfake}, and Arabic splits from MLAAD~\cite{muller2024mlaad}; and a \emph{multilingual pool} ($\approx$264k clips) adding SpoofCeleb~\cite{jung2024spoofceleb} and ASVspoof~5~\cite{wang2024asvspoof5}, which covers a range of codecs.
    \item \textbf{Data augmentation pipelines:} Combinations of DSU, RawBoost~\cite{tak2022rawboost}, MUSAN noise~\cite{snyder2015musan}, RIR reverberation~\cite{ko2017rir}, lossy compression (MP3/OGG), and speed perturbation ($0.9\text{--}1.1\times$). A specialized dialect model (DIA) applies speed perturbation and DSU without acoustic filtering.
\end{enumerate}

\subsection{Track-Adaptive Scoring and Fusion}
\label{sec:fusion}

\paragraph{Segment aggregation.}
For an input utterance $u$ and detector $i$, scoring uses either multi-window averaging or center cropping. \emph{Multi-segment scoring} divides $u$ into $K_u$ overlapping windows $\{c_1,\dots,c_{K_u}\}$ of 4.04\,s (stride 2.02\,s) and computes the mean score, providing stable estimates over variable-length dialectal recordings in Track~1. Conversely, \emph{center-crop scoring} evaluates a single 4\,s window, which performed better under Track~2 channel distortions where averaging across clean and distorted segments reduced performance.

\paragraph{Late score fusion.}
To handle score calibration differences across models, raw output scores $s_i$ are normalized via transductive $z$-score standardization over the evaluation set $\mathcal{E}$ and aggregated using non-negative fusion weights $\{w_i\}$:
\begin{equation}
z_i(u) = \frac{s_i(u) - \mu_i}{\sigma_i}, \qquad S(u) = \frac{\sum_i w_i z_i(u)}{\sum_i w_i},
\end{equation}
where $\mu_i$ and $\sigma_i$ denote the mean and standard deviation of detector $i$ over $\mathcal{E}$. This $z$-norm is transductive: a fused score depends on the rest of $\mathcal{E}$ and would not be reproduced by scoring files one at a time.

\paragraph{Track-adaptive weighting.}
The weight distribution differs between evaluation tracks. For \emph{Track~1} (dialect generalization), fusion uses a broad ensemble of decorrelated detectors, with complementary models (FTX, MEGA, FTXA) up-weighted ($w_i=1.5$). For \emph{Track~2} (acoustic robustness), performance is highest with a compact core ensemble of channel-matched detectors at $w_i=1.0$, with auxiliary models (FTX, MEGA, QWN) included at reduced weights ($w_i=0.5$). Including track-mismatched detectors at full weight increased Track~2 EER.

\section{Experimental Setup}
\label{sec:setup}

Audio clips are resampled to 16\,kHz and padded or cropped to 64{,}600 samples ($\approx$4.04\,s) during training. WPT models are trained for 8 epochs using AdamW~\cite{loshchilov2019adamw} ($\beta_1=0.9, \beta_2=0.999$, weight decay $10^{-4}$) with an initial learning rate of $10^{-4}$, a batch size of 8, and a 500-step linear warmup followed by cosine decay. Partially fine-tuned models (FTBB, FTX, FTXA) train for 4 epochs with learning rates of $5\times10^{-5}$ for the pooling head and $10^{-5}$ for unfrozen backbone layers. DSU perturbation probability is set between $0.5$ and $0.7$. Checkpoints are selected by minimum official-development EER, except CDEV (trained on train+dev), for which we use the final epoch. Fusion weights and segment aggregation were chosen after training from cached scores on the development set and the dev-test leaderboard, without final-test labels.

\section{Results and Discussion}
\label{sec:analysis}

\subsection{Official Submission Results}

On the official final tests, fusion of the detectors in Table~\ref{tab:detectors} yields \textbf{1.96\% EER} on Track~1 and \textbf{1.04\% EER} on Track~2, against XLS-R+AASIST baselines of 14.54\% and 27.11\% (87\% and 96\% relative reductions). On Track~1, a six-model W2V-BERT ensemble (T2, S123, S456, C123, C456, FTBB) scores 2.16\%; adding and up-weighting less correlated models trained on external data (FTX, MEGA, FTXA) lowers EER to 1.96\%. On Track~2, a compact core (FTBB, C456, T2) scores 1.06\% and reaches 1.04\% when FTX, MEGA, and QWN are added at $w{=}0.5$. Full-weight auxiliaries (1.10\%) or the dialect-only DIA model (1.09\%) raise EER.

\subsection{Per-Detector Cross-Split Analysis}
\label{sec:permodel}

Following the release of ground-truth test labels, we evaluate base detectors across challenge splits (Table~\ref{tab:detectors}).

First, score fusion outperforms every individual constituent detector across all test splits (Track~1 final: $3.01\% \rightarrow 1.96\%$; Track~2 final: $1.39\% \rightarrow 1.04\%$). 1,000-sample bootstrap resampling on the final-test splits yields non-overlapping 95\% confidence intervals for fusion vs.\ the best single model on Track~1 ($[1.88, 2.05]$ vs.\ $[2.90, 3.12]$) and Track~2 ($[0.97, 1.10]$ vs.\ $[1.33, 1.47]$).

Second, W2V-BERT models consistently outperform alternative backbones. XLS-R (5.84\%) and Dasheng (12.10\%) struggle under Track~1 dialect shift, while Qwen3-Omni reaches 3.03\% EER on Track~2 and is useful in that fusion. We hypothesize that Dasheng and Qwen3-Omni, both spectrogram frontends pre-trained on general audio, preserve channel cues more than dialectal structure, which matches their weak Track~1 EERs (12.10\%, 8.88\%) and stronger Track~2 EERs (3.60\%, 3.03\%).

Third, in-distribution \texttt{dev} EER ($0.04\text{--}1.37\%$) is not a proxy for out-of-distribution robustness: base-detector test EERs rise to $2.76\text{--}17.32\%$ ($5\text{--}200\times$ depending on model and shift).

\subsection{Asymmetric Fusion Dynamics}

Dialect shift and acoustic channel shift respond to different fusion strategies:
\begin{itemize}
    \setlength{\itemsep}{1pt}\setlength{\parsep}{0pt}\setlength{\parskip}{1pt}
    \item \textbf{Track 1 (Dialect shift):} Benefits from multi-window utterance averaging and an ensemble of decorrelated detectors. W2V-BERT models form correlated clusters ($r=0.85\text{--}0.97$), whereas models trained on external corpora (FTX, MEGA) yield less correlated predictions ($r=0.46\text{--}0.62$). Up-weighting these decorrelated candidates improves Track~1 performance.
    \item \textbf{Track 2 (Acoustic shift):} Benefits from center-crop scoring and a compact core ensemble anchored by channel-augmented models. Adding auxiliary detectors at full weight increases variance, whereas down-weighting them ($w=0.5$) maintains performance under channel distortions.
\end{itemize}

\section{Conclusion}
\label{sec:conclusion}

We described the UZH-CL systems for the ArA-DF 2026 Shared Task. Frozen W2V-BERT-2.0 with Wavelet Prompt Tuning, combined with track-specific score fusion, reached 1.96\% EER on Track~1 and 1.04\% EER on Track~2. Dialect shift and acoustic channel shift did not favor the same ensemble: a broader multi-window fusion helped Track~1, and a smaller channel-matched set with center-crop scoring helped Track~2.

\clearpage
\section{Limitations}
\label{sec:limitations}

Fusion uses transductive $z$-norm over the evaluation set, so a file's fused score depends on the rest of the batch; we do not report an inductive variant with development-set statistics. Each detector is a single training run, and close Track~1 EERs in Table~\ref{tab:detectors} (e.g., 3.01--3.15) should not be read as stable without seed variance. We did not deduplicate the external Arabic and multilingual pools against ArA-DF evaluation material by speaker, recording, or synthesis system, because those identities were not released under blind evaluation; the Track~1 fusion includes external-data models (FTX, MEGA, FTXA) and should be interpreted with that caveat. We also do not report the task's secondary metrics (Accuracy and macro-F1).

\section{Ethics Statement}
\label{sec:ethics}

This work is a defensive deepfake detector for Arabic speech. We use official ArA-DF data and publicly cited corpora under their licenses, and we do not release a generation system. Speech can identify speakers; we do not attempt re-identification. Reported EERs apply only to the shared-task conditions and are not a guarantee of unconstrained robustness.

\bibliography{custom}

\end{document}